\documentclass[12pt,a4paper]{article}
\usepackage{amsmath}
\usepackage{amssymb} 
\begin{document}
\begin{titlepage} 
\begin{flushright} IFUP--TH/2020\\ 
\end{flushright} ~
\vskip .8truecm 
\begin{center} 
\Large\bf Real analyticity of accessory parameters
\end{center}
\vskip 1.2truecm
\begin{center}
{Pietro Menotti} \\ 
{\small\it Dipartimento di Fisica, Universit{\`a} di Pisa}\\ 
{\small\it 
Largo B. Pontecorvo 3, I-56127, Pisa, Italy}\\
{\small\it e-mail: pietro.menotti@unipi.it}\\ 
\end{center} 
\vskip 0.8truecm
\centerline{February 2020}
                
\vskip 1.2truecm
                                                              
\begin{abstract}
We consider the problem of the real analytic dependence of the
accessory parameters of Liouville theory on the moduli of the problem,
for general elliptic singularities.  We give a simplified proof of the
almost everywhere real analyticity in the case of a single accessory
parameter as it occurs e.g. in the sphere topology with four sources
or for the torus topology with a single source by using only the
general analyticity properties of the solution of the auxiliary
equation.  We deal then the case of two accessory parameters. We use
the obtained result for a single accessory parameter to derive
rigorous properties of the projection of the problem on lower
dimensional planes. We derive the real analyticity result for two
accessory parameters under an assumption of irreducibility.
\end{abstract}

\end{titlepage}

\section{Introduction}

In several developments of Liouville theory it is necessary to exploit
the nature of the dependence of the accessory parameters on the moduli
of the problem. Typical field in which one exploits the real analytic
nature of the dependence of the accessory parameters on the moduli, is
the proof of the Polyakov relation, relating the accessory parameters
to the derivative of the on shell action of Liouville theory with
respect to the position of the singularity \cite{CMS1,CMS2,ZT,TZ} and
its extension \cite{torusblocks} to the moduli of an hyperelliptic
surface.

With regard to the proof of the real analytic dependence of the
accessory parameters on the position of the singularity we have, for
the case of parabolic and finite order singularities the
result by Kra \cite{kra} according to which the accessory parameters
are real analytic functions (not analytic functions) of the moduli
i.e. of the position of the singularities.

Parabolic singularities (sometimes called punctures) are characterized
by the strength of the source given by $\eta=1/2$ (see
eq.(\ref{liouville})), while finite order singularities are elliptic
singularities with source strength given by $\eta=(1-1/n)/2$, $n\in
Z_+$.

Kra employs, in presence of only parabolic and finite order
singularities, the possibility of using fuchsian mapping techniques.

Such technique is not available in presence of general elliptic
singularities.

In presence of a single accessory parameter, e.g. the sphere with the
four-sources or the torus with one source, it was proven in
\cite{torusIII,highergenus} that also for general elliptic
singularities, the accessory parameter is a real analytic function of
the moduli except for a zero measure region i.e. almost everywhere
(a.e.) in the moduli space. In the proof of such a result one exploits
only the uniqueness result for the value of the accessory parameter
combined with very general analytic properties of the matrix elements
of the monodromy matrices. Such analytic properties are a direct
consequence of the representation of the monodromy matrices in term of
the solutions of the auxiliary differential equation.

In the present paper we give a simplified proof of such a result and explore
the general features of the problem when more than one accessory
parameter is present, always using only the general analytic
properties of the solutions of the auxiliary equation.

To keep the formalism into reasonable complexity we consider the case
of two accessory parameters, which is the case e.g. of the sphere with
five sources or of the torus with two sources.

We give rigorous results on the projection of the problem on lower
dimensional planes which is a necessary step toward the analyticity
result. At this stage however it appears that the general analytic
properties of the solutions of the auxiliary equation are not
sufficient to progress. We show that under an irreducibility
assumption one reaches the final result of the a.e. real analyticity
also in presence of two accessory parameters. Such irreducibility
properties however should be derived by poking more deeply into the
consequences of auxiliary equation.

The paper is organized as follows: In section 2 we give the simplified
treatment of the one accessory parameter problem. In section
\ref{twoaccessory} we deal with the two accessory parameter problem
giving in subsection \ref{generalresult} the results which follow
from the general analytic properties of the solutions of the auxiliary
equation and in subsection 
\ref{irreduciblecase} the treatment of the irreducible case.  In
section \ref{conclusions} we summarize the obtained results and we
give a discussions of the possible developments. In the Appendix the
proof is given of a lemma which is instrumental for all the described
developments.

\section{The case of a single accessory parameter}\label{oneparameter}

We recall that the Liouville conformal field $\phi$ which satisfies
the partial differential equation
\begin{equation}\label{liouville}
-\partial_z\partial_{\bar z}\phi+e^\phi= 2\pi\sum_j \eta_j\delta^2(z-z_j)
\end{equation}
can be expressed in terms of the solutions of the auxiliary ordinary
differential equation in the complex plane
\begin{equation}\label{auxiliaryequation}
y''+Qy=0
\end{equation}
as
\begin{equation}\label{conffact}
e^{-\frac{\phi}{2}}= \frac{1}{\sqrt{2}|w_{12}|}\big[
\kappa^{-2} \overline{y_1(z)}y_1(z)-\kappa^{2} \overline{y_2(z)}y_2(z)\big]
\end{equation}
where $y_1,y_2$ are two independent solution of
(\ref{auxiliaryequation}) $w_{12}$ their wronskian and in the case of
the sphere topology
\begin{equation}
Q= \sum_j \frac{\eta_j(1-\eta_j)}{(z-z_j)^2}+\frac{\beta_j}{2(z-z_j)}~.
\end{equation}
The $\kappa$ is a real parameter.  In the case of the torus and all
hyperelliptic surfaces again the solution of eq.(\ref{liouville}) can
be reduced to the solution of a similar equation \cite{highergenus}.
The $\beta_j$ are
the accessory parameters which have to be chosen along with $\kappa$
as to have the $\phi$ appearing in eqs.(\ref{liouville},
\ref{conffact}) single valued.

The accessory parameters which realize the single valued solution of
eq.(\ref{liouville}) are unique. This is seen by recalling the
uniqueness and existence theorem for the solution of eq.(\ref{liouville})
\cite{picard,poincare,lichtenstein,troyanov,existence}
and noticing that
\begin{equation}\label{Qphi}
e^{\frac{\phi}{2}}\partial_z^2e^{-\frac{\phi}{2}}=
-Q(z)~~~~\beta_j=\frac{1}{i\pi} \oint Q(z)dz
\end{equation}
and thus each $\beta_j$ can be recovered from a contour integral
(\ref{Qphi}) in the $z$-plane. It is easily seen that
the $\beta_j$ are continuous functions of the moduli \cite{CMS1,CMS2}.

In addition the $\beta_j$ are subject to Fuchs conditions
\cite{torusIII,highergenus}: e.g. for the sphere with four sources we
have three $\beta$'s and two Fuchs conditions and thus one independent
accessory parameter to be determined and the same for the torus with
one source.  In the present section we shall refer for concreteness to
the sphere with four sources, but there is no difference in the
treatment.

For the location of the sources one can choose
$z_1=0,z_2=1,z_3=\infty$ and $z_4=u$ and after choosing a canonical
basis around the source at $z=0$ let $M(C_1)$ be the monodromy around
the source at $1$ and $M(C_2)$ the monodromy around the source at
$u$. After the imposition of single valuedness of the Liouville
field around $z=1$ and $z=u$, i.e.  imposition of the $SU(1,1)$ nature
of their monodromies we have that the $SU(1,1)$ nature of the
monodromy at infinity is also assured.

The monodromy matrix around a singularity is given by \cite{torusIII}
\begin{equation}\label{explicitM}
M=
\begin{pmatrix}
\tilde y_1&\tilde y'_1\\
\tilde y_2&\tilde y'_2
\end{pmatrix}
\begin{pmatrix}
y'_2&-y'_1\\
-y_2& y_1
\end{pmatrix}~.
\end{equation}
where $y_1,y_2$ are the solution at a point $z$ and
$\tilde y_1,\tilde y_2$ are the same solutions computed at the same
point $z$ after encircling the singularity.
The solutions $y_j$ at any point in the complex $z$-plane are
obtained from the absolutely convergent Cauchy series along a path
which keeps at a finite distance of the singularities and as such the
$y_j$ are analytic function of the parameters appearing in $Q$ and
thus of the free accessory parameter and of the position of the
singularity $z_4=u$.

Thus also the
matrix elements of all the monodromy matrices are analytic functions
of the free accessory parameter $\beta$ and of $u$.

From (\ref{conffact}) we see that $\kappa$ is given by the relative
weight of the two canonical solution $y_1,y_2$ at $z=0$ and thus is
fixed by the ratio of $M_{12}$ to $M_{21}$. $M(C_0)$ is diagonal and
as $\kappa^2$ is determined by the unique solution of the Liouville
equation (\ref{liouville}) we must have $M_{12}(C_1)\neq0$ or
$M_{12}(C_2)\neq0$. Let  $M_{12}(C_1)\neq0$ and thus also
$M_{21}(C_1)\neq0$.

Then we can the form the ratios
\begin{equation}
A(\beta,u) = \frac{M_{12}(C_2)}{M_{12}(C_1)}
\end{equation}
\begin{equation}
B(\beta,u) = \frac{M_{21}(C_2)}{M_{21}(C_1)}
\end{equation}
From eq.(\ref{explicitM}) the matrix elements of $M(C_j)$ are analytic
functions of the free accessory parameter $\beta$ and of the modulus
$u$.

The $SU(1,1)$ nature of the monodromy i.e. the possibility of choosing
a $\kappa$ such that all monodromies become $SU(1,1)$ implies
\begin{equation}\label{basiceq}
A(\beta,u) = \bar B(\bar\beta,\bar u).
\end{equation}
Once eq.(\ref{basiceq}) is satisfied, by exploiting the freedom on the
real parameter $\kappa$ we can obtain
$M_{21}(C_1)=\overline{M_{12}(C_1)}$ and
$M_{21}(C_2)=\overline{M_{12}(C_2)}$. On the other hand the monodromies
$M(C_j)$ are by construction $SL(2,C)$ and thus we have 
\begin{equation}
M_{11}(C_j)M_{22}(C_j)=1+M_{12}(C_j)M_{21}(C_j)>1~.
\end{equation}
Being the singularity elliptic we have
$M_{11}(C_j)+M_{22}(C_j)=-2\cos\alpha_j={\rm real}$, giving
$M(C_j)\in SU(1,1)$ which is the necessary and sufficient condition
for having a single valued $\phi$. We recall now that $M(C_0)$ is
already $SU(1,1)$, $M(C_1)$ and $M(C_2)$ become $SU(1,1)$ using
eq.(\ref{basiceq}) and $M(C_\infty)$ becomes $SU(1,1)$ as a
consequence.

Thus the satisfaction of (\ref{basiceq}) is both necessary and
sufficient for the single-valuedness of $\phi$ all over the plane.

It is useful now to proceed with the method of polarization i.e. to
consider $\beta$ and $\bar\beta$, and $u$ and $\bar u$ as independent
complex variables. Then eq.(\ref{basiceq}) becomes the system
\begin{eqnarray}\label{SYS1}
&&A(\beta,u) = \bar B(\beta^c,u^c)\nonumber\\
&&B(\beta,u) = \bar A(\beta^c,u^c)
\end{eqnarray}
with the proviso that at the end we shall be interested into the
self-conjugate (s.c.) solutions of eq.(\ref{SYS1}) i.e. in the
solution for which at $u^c=\bar u$ we have $\beta^c=\bar\beta$.

We are interested in the analytic behavior of $\beta$ in a
neighborhood of $u_0$ and to simplify the notation we shall from now
on denote with $u$ the difference $u-u_0$ and with $\beta$ the
difference $\beta-\beta(u_0)$. In addition we shall denote with
$A(\beta,u)$ the difference $A(\beta,u)-A(\beta(u_0),u_0)$ and the
same for $B$ so that now $A(0,0)=0,~B(0,0)=0$.

We notice that due to the structure of (\ref{SYS1}) if $\beta,\beta^c$
is a solution at $u,u^c$, $\bar\beta^c,\bar\beta$ is a solution at
$\bar u^c,\bar u$ in particular if the solution of (\ref{SYS1}) at
$u=u^c=0$ is unique it is self-conjugate and thus due to uniqueness of
the physical solution $\beta=\beta^c=0$.

Before entering the details we outline the structure of the proof of
the real analyticity of $\beta(u)$.  If for $u=u^c=0$ the origin
$\beta=\beta^c=0$ is an isolated solution of the polarized system
(\ref{SYS1}), then as we shall see the real analyticity of
$\beta$ easily follows. What we shall prove is that if the origin
$\beta=\beta^c=0$ is not an isolated solution then we can construct
a solution with $\beta^c=\bar\beta\neq0$ thus violating the uniqueness
theorem.

We come now to the details.
$A(\beta,0)$ and/or  $B(\beta,0)$ depend on $\beta$ otherwise the
system (\ref{basiceq}) would not determine $\beta=\beta^c=0$ for $u=u^c=0$.
 
Using Weierstrass preparation theorem (WPT)
\cite{whitney,gunningrossi,bochnermartin} we can rewrite system
(\ref{SYS1}) as
 
\begin{eqnarray}\label{twoWpol}
&&P_1(\beta^c|\beta,u,u^c)=0\nonumber\\
&&P_2(\beta^c|\beta,u,u^c)=0~.
\end{eqnarray}
 
Necessary and sufficient condition for the two polynomials
(\ref{twoWpol}) to have a common root $\beta^c$ is the vanishing of
their resultant  \cite{whitney,harris}
\begin{equation}\label{resultant}
R(P_1,P_2)\equiv h(\beta,u,u^c)=0~.
\end{equation}
In particular from the existence result we know
\begin{equation}
h(0,0,0)=0.
\end{equation}
We exploit again Weierstrass preparation theorem applied to
(\ref{resultant}) writing
\begin{equation}\label{hequation}
h(\beta,u,u^c)=w(\beta,u,u^c) P(\beta|u,u^c)=0
\end{equation}
being $w(\beta,u,u^c)$ a unit, i.e. an analytic function non zero in a
neighborhood of $\beta=u=u^c=0$. However to apply the WPT to $h$ we
must have that $h(\beta,0,0)\not\equiv0$. We shall prove that if
$h(\beta,0,0)\equiv0$, which means that we can solve eq.(\ref{SYS1})
for $u=u^c=0$ for any $\beta$ in a neighborhood of zero, then we have
s.c. solutions of eq.(\ref{SYS1}) at $u=u^c=0$ for $\beta\neq0$
different from zero, which goes against the uniqueness
theorem. Reducing the dependence of $\beta$ on $u,u^c$ through the non
trivial relation $P(\beta|u,u^c)=0$ is the major step in analyzing the
analytic properties of such a dependence \cite{torusIII}.

\bigskip

As we mentioned above we prove now that if $h(\beta|0,0)\equiv0$ then
we have s.c. solutions of eq.(\ref{SYS1}) at $u=u^c=0$ for $\beta$
different from zero thus violating the uniqueness result.

\bigskip

We consider first the case in which the analytic function
of $\beta$, $A(\beta,0)$ is of order $1$
at $\beta=0$ i.e. $A(\beta,0) = \beta~a(\beta,0)$ with $a(0,0)\neq0$.
Then the system (\ref{SYS1}) goes over to
\begin{eqnarray}
&&\beta a(\beta,0) = \beta^c \bar b(\beta^c,0)\label{SYS1b1}\\
&&\beta b(\beta,0) = \beta^c \bar a(\beta^c,0)\label{SYS1b2}~.
\end{eqnarray}
Multiplying we obtain for the solutions of the system
\begin{equation}\label{ababbb} 
a(\beta,0)\bar a(\beta^c,0)=b(\beta,0)\bar b(\beta^c,0)~.
\end{equation} 
We notice that eq.(\ref{SYS1b1}) given
$\beta$ near $\beta=0$ can be solved in the form of the convergent series
\begin{equation}\label{series} 
\beta^c = \frac{a(0,0)}{\bar b(0,0)} \beta + c_2 \beta^2 +\dots
\end{equation}
and similarly for eq.(\ref{SYS1b2}) given $\beta^c$.
This is the result of the implicit function theorem
\cite{whitney,gunningrossi}. In the limit $\beta\rightarrow0$ we
have
\begin{equation}\label{bcondition} 
\frac{\beta^c}{\beta}\rightarrow \frac{a(0,0)}{\bar b(0,0)}=
\frac{b(0,0)}{\bar a(0,0)}~.  
\end{equation} 

An other consequence of the system (\ref{SYS1b1},\ref{SYS1b2}) is
\begin{equation}\label{oureq} 
\beta^2 a(\beta,0) b(\beta,0)=({\beta^c})^2 \bar a(\beta^c,0) 
\bar b(\beta^c,0)~.
\end{equation} 
We look for a s.c. solution of (\ref{oureq}) i.e. a solution with
$\beta^c=\bar\beta$ or more explicitly $\beta=\rho e^{i\alpha(\rho)}$,
$\beta^c=\rho e^{-i\alpha(\rho)}$ where the unknown is $\alpha(\rho)$,
with boundary condition (\ref{bcondition}).

The function $\alpha(\rho)$ is given by the vanishing of the real
function
\begin{equation}
f(\rho,\alpha)=i(e^{2i\alpha(\rho)}a(\rho e^{i\alpha(\rho)},0) b(\rho
e^{i\alpha(\rho)},0)- e^{-2i\alpha(\rho)} \bar a(\rho
e^{-i\alpha(\rho)},0) \bar b(\rho e^{-i\alpha(\rho)},0))~. 
\end{equation}   
For solving the equation $f(\rho,\alpha)=0$ in the neighborhood of
$\rho=0$ we use the real implicit function theorem. We have from
(\ref{bcondition},\ref{ababbb})
\begin{equation}
f(0,\alpha(0,0)))=0
\end{equation}
and
\begin{equation}
\frac{\partial
  f(\rho,\alpha)}{\partial\alpha}\bigg|_{\rho=0}=
-2 (e^{2i\alpha(0)}a(0,0)b(0,0)+ e^{-2i\alpha(0)}\bar a(0,0)\bar
b(0,0))=
-4 e^{2i\alpha(0)}a(0,0)b(0,0)\neq 0~.
\end{equation}
Thus chosen $\alpha(0)$ satisfying (\ref{bcondition}), $\alpha(\rho)$
exists and is unique around $\rho=0$.  We show now that $\beta=\rho
e^{i\alpha(\rho)}$, $\beta^c=\rho e^{-i\alpha(\rho)}$ is a solution of
the system (\ref{SYS1b1},\ref{SYS1b2}).  Always working in a
neighborhood of the origin we know that given any $\beta$, and in
particular $\beta=e^{i\alpha(\rho)}\rho$, we have a unique $\beta^c$
which solves (\ref{SYS1b1},\ref{SYS1b2}). On the other hand given $\beta$, the
solutions $\beta^c$ of eq.(\ref{oureq}) with boundary conditions
$\beta^c/\beta = a(0,0)/\bar b(0,0)$  is unique.  In fact setting
$\beta^c=\Omega\bar\beta$, $\Omega(0)=1$ to fulfill the boundary
conditions we have with
\begin{equation}\label{Fequation}
F=\Omega^2\bar a(\Omega\bar\beta,0)\bar b(\Omega\bar\beta,0)-
\bar a(\bar\beta,0)\bar b(\bar\beta,0)
\end{equation}
\begin{equation}
F(0) =0,~~~~~\frac{\partial F}{\partial\Omega}\bigg|_{\Omega=1,\bar\beta=0}
=2 \bar a(0,0)\bar b(0,0)\neq0~
\end{equation}
and using the implicit function theorem we have that $\Omega\equiv1$ is the
unique solution of eq.(\ref{Fequation}) in a neighborhood of $\rho=0$.
Thus being the solution of eq.(\ref{oureq}) for $\beta=\rho e^{i\alpha(\rho)}$ 
unique, this is also the unique solution of eq.(\ref{SYS1}) at
$u=u^c=0$.

In conclusion we found a non zero s.c. to the original system
(\ref{SYS1}) at $u=u^c=0$ and this goes against the uniqueness theorem
for the solution of the Liouville equation which implies the
uniqueness of the accessory parameters.

\bigskip

We treat now the case in which $A$ and $B$ are of higher order at the
origin. We have
\begin{eqnarray}
&&\beta^m a(\beta,0)=(\beta^c)^n\bar b(\beta^c,0)\label{1betamI}\\
&&  
\beta^n b(\beta,0)=(\beta^c)^m\bar a(\beta^c,0)\label{1betamII}~.
\end{eqnarray}
Then $n=m$. We have also for a solution of eqs.(\ref{1betamI},
\ref{1betamII})
the equation
\begin{equation}\label{ourmequation}
\beta^{2m} a(\beta,0)b(\beta,0)=
(\beta^c)^{2m}\bar a(\beta^c,0)\bar b(\beta^c,0)~.
\end{equation}
and also for $\beta\rightarrow0$
\begin{equation}\label{kroot}
\bigg(\frac{\beta^c}{\beta}\bigg)^m\rightarrow 
\frac{a(0,0)}{\bar b(0,0)}=\frac{b(0,0)}{\bar a(0,0)}~.
\end{equation}
The root of eq.(\ref{kroot}) is not a choice but is given by the
hypothesis of existence of a solution, which we want to
disprove. Around such a root
\begin{equation}
\bigg(\frac{a(\beta,0)}{\bar b(\beta^c,0)}\bigg)^{\frac{1}{m}}
\end{equation}
is an analytic function both of $\beta$ and $\beta^c$, being $a$ and $b$
units and thus under such boundary condition $\beta^c$ is the unique
solution of the system (\ref{1betamI}, \ref{1betamII}).

Again we solve as previously the equation
\begin{equation}
\beta^{2m}a(\beta,0)b(\beta,0)=(\bar\beta)^{2m}\bar a(\bar\beta,0)
\bar b(\bar\beta,0)
\end{equation}
with $\beta=\rho e^{i\alpha(\rho)}$, $\bar\beta=\rho
e^{-i\alpha(\rho)}$ and with 
\begin{equation}
e^{-2i\alpha(0)}=\bigg(\frac{a(0,0)}{\bar b(0,0)}\bigg)^{\frac{1}{m}}~.
\end{equation}

Again given $\beta=\rho e^{i\alpha(\rho)}$ the
solution of eq.(\ref{ourmequation}) is unique and thus the unique
solution of the system (\ref{1betamI}, \ref{1betamII}) with our
$\alpha(\rho)$ is given by $\beta^c=\rho e^{-i\alpha(\rho)}$ and
we obtained a non zero s.c. solution of the initial system (\ref{SYS1})
thus violating the uniqueness theorem.

We have reached the conclusion that $h(\beta,0,0)\not\equiv0$ and thus
we can apply the WPT giving $\beta$ as solution of $P(\beta|u,u^c)=0$,
with the $P$ appearing in (\ref{hequation}).

For completeness we recall the proof of how from
$h(\beta|0,0)\not\equiv0$ the a.e. real analyticity of $\beta$
follows.

From $h(\beta|0,0)\not\equiv0$ we derived the equivalent
W-polynomial $P(\beta|u,u^c)$.
We notice that as a rule
$P(\beta|u,u^c)=0$ in addition to the physical solution
$\beta(u,\bar u)$ will have other solutions for $u^c\neq \bar
u$. We decompose $P$ in irreducible components
\begin{equation}
P(\beta|u,u^c)=P_1(\beta|u,u^c)\dots P_n(\beta|u,u^c)~.
\end{equation}
We know that the discriminant $D_j(u,u^c)$ of $P_j$ is an analytic
function of $u,u^c$ which is not identically zero and thus vanishes on
a {\it thin} \cite{whitney} set $\cal S$ of $u,u^c$ which as such is
of zero 4-dimensional measure in $u,u^c$ \cite{gunningrossi}. Then we
have that except for such a set all the solutions of $P_j=0$ are
distinct and thus analytic function of $u,u^c$ \cite{whitney} and in
particular the physical solution, obtained setting $u^c=\bar u$,
is a real analytic function of $u$.  The subset ${\cal S}'$ of
$\cal S$ with $u^c=\bar u$ has zero 2-dimensional measure
\cite{highergenus}.

\section{The case of two accessory parameters}\label{twoaccessory}

\subsection{General results}\label{generalresult}
 
We come now to the more complicated case of two accessory
parameters. Typical examples are the sphere with five sources and the
torus with two sources. This time we need three monodromies to
determine the $\beta_1$ and $\beta_2$. We shall be interested in the
dependence of the accessory parameters on a single modulus which we
shall call $u$ , e.g.  the position of the third source in the problem
of the sphere with five sources as the same treatment can be repeated
for the other moduli. Moreover after a shift we shall work in the
neighborhood of $u=0$. This covers the general case.

Following the discussion given at the beginning of the previous
section we define, with $M_{12}(C_1)\neq0$ and thus $M_{21}(C_1)\neq0$
\begin{equation}
A(\beta_1,\beta_2,u) = \frac{M_{12}(C_2)}{M_{12}(C_1)}
\end{equation}

\begin{equation}
B(\beta_1,\beta_2,u) = \frac{M_{21}(C_2)}{M_{21}(C_1)}
\end{equation}

\begin{equation}
C(\beta_1,\beta_2,u) = \frac{M_{12}(C_3)}{M_{12}(C_1)}
\end{equation}

\begin{equation}
D(\beta_1,\beta_2,u) = \frac{M_{21}(C_3)}{M_{21}(C_1)}~.
\end{equation}

The $SU(1,1)$ relations for the monodromy matrices, after performing
polarization, form two systems of equations and with the already
discussed shift in the $\beta_j$ the $u$ and the $A,B,C,D$, they
become
\begin{eqnarray}\label{SYS2}
&&A(\beta_1,\beta_2,u)=\bar B(\beta^c_1,\beta^c_2,u^c)~~~~~~~~I\nonumber\\
  &&B(\beta_1,\beta_2,u)=\bar A(\beta^c_1,\beta^c_2,u^c)~~~~~~~~I^c\nonumber\\
~\\
&&C(\beta_1,\beta_2,u)=\bar D(\beta^c_1,\beta^c_2,u^c)~~~~~~~~II\nonumber\\
&&D(\beta_1,\beta_2,u)=\bar C(\beta^c_1,\beta^c_2,u^c)~~~~~~~~II^c.\nonumber
\end{eqnarray}
with $A(0,0,0)=0$ and the same for $B,C,D$.
We notice again the self-conjugate structure of the system (\ref{SYS2}):
If at $u,u^c$, $\beta_1,\beta_1^c,\beta_2,\beta_2^c$ is a solution of 
(\ref{SYS2}) then $\bar\beta^c_1,\bar{\beta_1},\bar\beta_2^c,\bar\beta_2$
is a solution of (\ref{SYS2}) at $\bar u^c,\bar u$.

The uniqueness theorem on the solution of Liouville equation tells us
that the unique s.c. solution to (\ref{SYS2}) at $u=u^c=0$ i.e. to
\begin{eqnarray}\label{SYS2orig}
&&A(\beta_1,\beta_2,0)=\bar B(\bar\beta_1,\bar\beta_2,0)\nonumber\\
&&C(\beta_1,\beta_2,0)=\bar D(\bar\beta_1,\bar\beta_2,0)
\end{eqnarray}
is $\beta_1=\beta_2=0$.
It will be useful, by performing a linear invertible transformation
\begin{equation}
\beta_j\rightarrow a_{j1}\beta_1+a_{j2}\beta_2,~~~~
\beta^c_j\rightarrow \bar a_{j1}\beta^c_1+\bar a_{j2}\beta_2^c~,
\end{equation}
to render $A(\beta_1,\beta_2,0)$ regular in both variables and the
same for $B$, $C$, $D$ by a single transformation \cite{bochnermartin}
without loosing the s.c. property of the system.  It means that if the
order of $A(\beta_1,\beta_2,0)$ is e.g. $2$ than after the
transformation $\beta_1^2$ appears with coefficient
$a_1(\beta_1,\beta_2)$ with $a_1(0,0)$  different from zero and
the same for $\beta_2$, i.e. $a_2(0,0)\neq0$.

We want to deal here with the general situation when the order of the
system $I,I^c$ and $II,II^c$ at $u=u^c=0$ is arbitrary.

Given the $I,I^c,II,II^c$ at $u=u^c=0$ if they all do not depend on
the $\beta$'s we can set $\beta_1^c=\bar\beta_1\neq0$,
$\beta_2^c=\bar\beta_2=0$ to have a violation of the uniqueness theorem.
From the structure of the equations we see that if the system depends
on $\beta_1$ it depends also on $\beta_1^c$. If at this point it does
not depend on $\beta_2$ and $\beta_2^c$ we can set
$\beta_1^c=\bar\beta_1=0$ and $\beta_2^c=\bar\beta_2\neq0$ to have a
violation of the uniqueness theorem. Thus (\ref{SYS2}) depends on all
the $\beta$'s. As we chose a regular set of variable we can eliminate
$\beta_2^c$. In fact as $\bar B(0,\beta_2^c,0)$ depends
on $\beta_2^c$ we have for some $\varepsilon$
\begin{equation} 
|\bar B(0,\beta_2^c,0)| >\eta ~~~~{\rm for}~~~~|\beta_2^c|=\varepsilon
\end{equation} 
and then
\begin{equation} 
|A(\beta_1,\beta_2,u)-\bar B(\beta_1^c,\beta_2^c,u^c)|>\eta/2
\end{equation} 
for $\Delta \times\{|\beta_2^c|=\varepsilon\}$ where
$\Delta=\{|\beta_1|^2+|\beta_1^c|^2+|\beta_2|^2+|u|^2+|u^c|^2<\varepsilon_1\}$.
This is a sufficient condition for the projection of the system
(\ref{SYS2}) on the hyperplane $\beta_1,\beta_1^c,\beta_2,u,u^c$
\cite{whitney}.  The projection is an analytic variety given by the
vanishing of a finite set of analytic functions \cite{whitney}
\begin{equation}\label{firstprojection} 
f_j(\beta_1,\beta_1^c,\beta_2,u,u^c)=0~.
\end{equation} 
Their number is given by
$N=(m+3)(m+2)(m+1)/3!-1$ \cite{whitney}, being $m$ the order of 
the Weierstrass polynomial for
$A(\beta_1,\beta_2,u)- \bar B(\beta_1^c,\beta_2^c,u^c)$ related to
the variable $\beta_2^c$. For $m=1$ such a number is $3$ as expected.
The number $N$ may depend on the chosen domain but we shall be
interested in the germ \cite{whitney,gunningrossi}
of the variety i.e. in an arbitrary non zero neighborhood of the origin.

It is important to recall that the vanishing of the functions
(\ref{firstprojection}) is both a necessary and sufficient condition 
for having above such a point a solution of the system 
(\ref{SYS2}). Obviously $f_j(0,0,0,0,0)=0$.

We now investigate the nature of the
$f_j(\beta_1,\beta_1^c,\beta_2,0,0)$. In the remainder of this section we
shall be interested in the variety at $V$ at $u=u^c=0$ which we
call $V_0$ and thus we shall omit the last argument in $A,B,C,D$
understanding that $u=u^c=0$  and the same in the $f_j$.

If all $f_j(0,0,\beta_2)$ do not depend on $\beta_2$ we have for
any $\beta_2$ and $\beta_1^c=\beta_1=0$ at least a $\beta_2^c$ which
solves system (\ref{SYS2}). 
In particular for any $\beta_2$ exists at least a $\beta_2^c$ 
which solves 
\begin{eqnarray}\label{reducedSYS} 
&&A(0,\beta_2)=\bar B(0,\beta^c_2)\nonumber\\
&&B(0,\beta_2)=\bar A(0,\beta^c_2)
\end{eqnarray} 
which is the one-accessory parameter problem we have already solved in
section \ref{oneparameter}. From the projection theorem we know that
given $\beta_2$ we have at least one solution of (\ref{reducedSYS})
which is also a solution of the the same system with $C,D$ replacing
$A,B$. Then following the procedure of section \ref{oneparameter} we
reach a s.c. solution i.e.  a solution with $\beta_2^c=\bar\beta_2$
violating the uniqueness theorem.

The conclusion is that $f_j(0,0,\beta_2)\not\equiv0$ i.e. for some
$j$ the $f_j$ is not identically zero.

We consider now the dependence of $f_j(0,\beta_1^c,0)$ on
$\beta_1^c$. If $f_j(0,\beta_1^c,0)\equiv0$ then we have solutions
for $\beta_1=\beta_2=0$ and any $\beta_1^c$. The consequence is that
for any $\beta_1^c$ we have at least a $\beta_2^c$ such that
\begin{eqnarray}\label{zerosystem}
&&A(0,0)=0=\bar B(\beta_1^c,\beta_2^c)\nonumber\\
&&B(0,0)=0=\bar A(\beta_1^c,\beta_2^c)\nonumber\\
&&C(0,0)=0=\bar D(\beta_1^c,\beta_2^c)\nonumber\\
&&D(0,0)=0=\bar C(\beta_1^c,\beta_2^c)
\end{eqnarray}
or for any $x$ we have a $y$ such that $A(x,y)=B(x,y)=C(x,y)=D(x,y)=0$
giving rise to a non zero solution of
\begin{eqnarray}
&&A(x,y)=\bar B(\bar x,\bar y)\nonumber\\
&&C(x,y)=\bar D(\bar x,\bar y)
\end{eqnarray}
violating again the uniqueness theorem.

Thus we reach the conclusion that $f_j(0,\beta_1^c,0)\not\equiv0$.

Using $f_j(0,0,\beta_2)\not\equiv0$ we can project out $\beta_2$ and
thus reach the projected analytic variety
$g^{(1)}_j(\beta_1,\beta_1^c)=0$ which are the necessary and
sufficient condition for having a solution of the system
$f_j(\beta_1,\beta_1^c,\beta_2)=0$ and thus a point of the variety
$V_0$ above $\beta_1,\beta_1^c$. Notice that due to the
s.c. structure of (\ref{SYS2}) we have also $\bar
g_j^{(1)}(\beta_1^c,\beta_1)=0$.  We could also have
$g^{(1)}_j(\beta_1,\beta_1^c)\equiv 0$ which means that the projection
of the variety on the plane $\beta_1,\beta_1^c$ is a whole open
neighborhood of $\beta_1=\beta_1^c=0$.

Similarly using $f_j(0,\beta_1^c,0)\not\equiv0$ we can project out
the variable $\beta_1^c$ and reach the projected analytic variety
$k_j(\beta_1,\beta_2)=0$ which is the necessary and sufficient
condition for having a point of the variety
$f_j(\beta_1,\beta_1^c,\beta_2)=0$ above $\beta_1,\beta_2$ and thus a
point of the original variety $V_0$ above $\beta_1,\beta_2$.

We further remark that if $\beta_1,\beta_1^c,\beta_2,\beta_2^c$ is a
point of $V_0$ then both $g^{(1)}_j(\beta_1,\beta_1^c)=0$ and
$k_j(\beta_1\beta_2)=0$ have to be satisfied. The reverse however is
not necessarily true, i.e.  if $\beta_1,\beta_1^c,\beta_2$ are such
that $k_j(\beta_1,\beta_2)=0$ and $g_j^{(1)}(\beta_1,\beta_1^c)=0$ is not
granted that above $\beta_1,\beta_1^c,\beta_2$ we have a point of the
variety because $g_j^{(1)}(\beta_1,\beta_1^c)=0$ does not assure that
above $\beta_1,\beta_1^c$ we have a point of the variety with the
chosen $\beta_2$.

All these constraints $g_j^{(k)}(\beta_k,\beta_k^c)=0$,
$k_j(\beta_1,\beta_2)=0$, $k_j^c(\beta_1^c,\beta_2^c)=0$ are necessary
and sufficient conditions for having above the given a pair of
$\beta$'s at least one solution of (\ref{SYS2}) at $u=u^c=0$~.

We remark that if $g^{(1)}_j(\beta_1,\beta_1^c)\not\equiv0$ then also
$k_j(\beta_1,\beta_2)\not\equiv0$. In fact if
$k_j(\beta_1,\beta_2)\equiv0$ we have a point the variety above
$\beta_1=0$ and any $\beta_2$. But below such point we must have
$g^{(1)}_j(0,\beta_1^c)=0$ i.e. $\beta_1^c=0$ if
$g^{(1)}_j(\beta_1,\beta_1^c)\not\equiv0$. Thus we are reduced to the
problem $\beta_1=\beta_1^c=0$ with the assurance that we have a
solution of the system (\ref{SYS2}) for any $\beta_2$. This is the
one-$\beta$ problem that we have already solved.

Vice-versa if $k_j(\beta_1,\beta_2)\not\equiv0$ and
$g^{(1)}_j(\beta_1,\beta_1^c)\equiv0$, for $\beta_1=0$ and any
$\beta_1^c$ we have a point on the variety $V_0$. But below
such point we must have also $\beta_2=0$ due to
$k_j(\beta_1,\beta_2)\not\equiv0$, and thus solutions with
$\beta_1=\beta_2=0$ and any $\beta_1^c$.  Explicitly for any
$\beta_1^c$ we have a $\beta_2^c$ which satisfies the system
(\ref{zerosystem})
and thus we have 
\begin{eqnarray}
&&0=A(\bar\beta_1^c,\bar\beta_2^c)=\bar B(\beta_1^c,\beta_2^c)\\
&&0=C(\bar\beta_1^c,\bar\beta_2^c)=\bar D(\beta_1^c,\beta_2^c)
\end{eqnarray}
which is a s.c. solution of the system (\ref{SYS2}).

We conclude that the relations $g_j^{(1)}(\beta_1,\beta_1^c)\not\equiv0$
and $k_j(\beta_1,\beta_2)\not\equiv0$ are equivalent.

We can perform the same reasoning eliminating the variable $\beta_2$
to reach a variety given by
\begin{equation}\label{beta2c}
f_j^c(\beta_1,\beta_1^c,\beta_2^c)=\bar f_j(\beta_1^c,\beta_1,\beta_2^c)=0
\end{equation}
with the result 
\begin{equation}
f_j^c(0,0,\beta_2^c)\not\equiv 0~~~~
{\rm  and}~~~~f_j^c(\beta_1,0,0)\not\equiv0
\end{equation}
from which the projections $g_j^{(1)}(\beta_1,\beta_1^c)=0$ and
$k_j^c(\beta_1^c,\beta_2^c)=0$ can be performed.

Moreover due to the regular choice of variables there is no
qualitative distinction between the index $1$ and $2$ we have also the
projections
\begin{equation}
g^{(2)}_j(\beta_2,\beta_2^c)=0~.
\end{equation}

We have reached the result that the conditions which express the
non trivial nature of the related projections
\begin{equation}\label{equivalentprojections}
g^{(1)}_j(\beta_1,\beta_1^c)\not\equiv0,~~~~
k_j(\beta_1,\beta_2)\not\equiv0,~~~~
k^c_j(\beta^c_1,\beta^c_2)\not\equiv0,~~~~ 
g^{(2)}_j(\beta_2,\beta_2^c)\not\equiv0
\end{equation}
are all equivalent. Furthermore we notice that due to the s.c. nature of 
the system (\ref{SYS2}) we have for the solutions of
$g_j^{(1)}(\beta_1,\beta_1^c)=0$ the validity of 
$\bar g^{(1)}(\bar\beta_1^c,\bar\beta_1)=0$ and the same with $1$
replaced by $2$ and  $k^c_j(\beta^c_1,\beta^c_2)=
\bar k_j(\beta^c_1,\beta^c_2)$. From now on we shall drop the upper
index $(1)$ in $g_j^{(1)}$. 

\bigskip

1) Let us consider first the case in which
$g_j(\beta_1,\beta_1^c)\not\equiv0$ with the consequences of
eq.(\ref{equivalentprojections}). In this case $\beta_1^c$ is driven
by $\beta_1$ and we have a variety of complex dimension $1$. We recall
that $f_j(0,0,\beta_2)\not\equiv0$. Then a general results
\cite{whitney} tells us that the variety $V_0$ is given by a union of
branches of a W-type variety i.e. of
\begin{eqnarray}\label{Wtype1}
&&P_1(\beta_1^c|\beta_1)=0\label{Pg} \\
&&P_2(\beta_2|\beta_1)=0\\
&&P_2^c(\beta_2^c|\beta_1)=0
\end{eqnarray}
where the $P$'s are Weierstrass polynomial with discriminant not
identically zero, thus W-polynomials with simple sheets.
We can also write
\begin{eqnarray}\label{1dimrepresentation}
&&P(\beta_2|\beta_1)=0\\
&&P^c(\beta_2^c|\beta_1^c)=0
\end{eqnarray}
where in the last the $\beta_1^c$ is driven by (\ref{Pg}).

We recall that due to the s.c. structure we have also the validity of
\begin{eqnarray}
&&\bar P_1(\beta_1|\beta_1^c)=0\\
&&\bar P_2(\beta_2^c|\beta_1^c)=0~.
\end{eqnarray}

In the present case i.e. $g_j(\beta_1,\beta_1^c)\not\equiv0$ we can
proceed to the further projection $h_l(\beta_1|u,u^c)=0$ and if
$h_l(\beta_1|0,0)\not\equiv0$ we can apply the procedure of
\cite{torusIII,highergenus} summarized at the end of section
\ref{oneparameter} of the present paper to prove that $\beta_1$ is a
real analytic function of $u$.

\bigskip

2) Let us consider now the case in which
$g_j(\beta_1,\beta_1^c)\equiv0$

The result $f_j(0,0,\beta_2)\not\equiv 0$ combined with the absence
of constraints between $\beta_1$ and $\beta_1^c$ makes $V_0$ a variety
of complex dimension $2$
and thus given by the union of a number of branches of the W-type
variety
\begin{eqnarray}
&&P_2(\beta_2|\beta_1,\beta_1^c)=0 \\
&&\bar P_2(\beta_2^c|\beta_1^c,\beta_1)=0~.
\end{eqnarray}

\bigskip

We notice that if we prove that $(0,0,0,0)$ is an isolated solution of
(\ref{SYS2}) at $u=u^c=0$, in the sense that for some $\varepsilon$
the origin is the only solution with $|\beta_j|<\varepsilon$,
$|\beta_j^c|<\varepsilon$ then we have both
$g_j(\beta_1,\beta_1^c)\not\equiv0$ and $h_l(\beta_1|0,0)\not\equiv0$
and this is enough to reach the a.e. real analyticity of $\beta_1$ as
a function of $u$.

In fact choosing as domain $H=\{|\beta_j|<\varepsilon$,
$|\beta_j^c|<\varepsilon\}$ the solutions of $h_l(\beta_1|0,0)=0$ are
exactly the projection of the solutions of the system (\ref{SYS2})
lying in $H$ and thus if there are no solutions in $H$ except the
origin, the only projection is $\beta_1=0$, i.e. the only solution of
$h_l(\beta_1|0,0)=0$ is $\beta_1=0$.  We recall that the projections
on lower dimensional planes and thus the $g_j$ and $h_l$ depend on the
chosen initial domain and we are concerned with a neighborhood of the
origin.

Summarizing, in this section we proved that we can always
project the variety $V_0$ on the planes $(\beta_1,\beta_1^c)$,
$(\beta_2,\beta_2^c)$, $(\beta_1,\beta_2)$, $(\beta_1^c,\beta_2^c)$.
These projections are all non trivial if any of the relations of
eq.(\ref{equivalentprojections}) are satisfied. If the further
projection $h_l(\beta_1|u,u^c)=0$ is non trivial we have that the
$\beta_1$ is a real analytic function of $u$. Furthermore such a
result is always true if one proves that the origin is an isolated 
point of $V_0$.

The above scheme worked perfectly in the case of one accessory
parameter; it is however too general for the two parameter case, in
the sense that we cannot prove, without further information, that the
projection $h_l(\beta_1|0,0)$ is non trivial. Such information should be
provided by a more detailed exploitation of the consequences of the
auxiliary equation (\ref{auxiliaryequation}) or even from
(\ref{liouville}). Here below as an illustration we deal with the
irreducible case.

\subsection{The irreducible case}\label{irreduciblecase}

As we saw in the previous subsection we have to examine the two
possibilities $g_j\not\equiv0$ and $g_j\equiv0$.

\bigskip

1. $g_j\not\equiv0$.

We shall now work under the assumption that the variety
$V_0$ is irreducible.

The variety $g_j=0$ is the projection of $V_0$ on the plane
$\beta_1,\beta_1^c$ and due to the irreducibility of $V_0$ the
projection $g_j=0$
is irreducible. From the properties of (\ref{equivalentprojections})
we see that in this case the dimension of $V_0$, i.e. the dimension of
the associated manifold $V_0^-$ is $1$.

\bigskip

Then the variety $V_0$ is described
by a union of branches of the W-type variety \cite{whitney}
\begin{eqnarray}\label{dimension1}
  &&P_g(\beta_1^c|\beta_1)=0\label{Pg1}\\
  &&P_2(\beta_2|\beta_1)=0\\
  &&P_2^c(\beta_2^c|\beta_1)=0 ~. 
\end{eqnarray} 
The $P_g,P_2,P^c_2$ are irreducible as a consequence of the
irreducibility of $V_0$; here $P_g$ plays the role of the $g$ of
the Appendix.

The system (\ref{dimension1}) can also be rewritten as
\begin{eqnarray}
  &&P_2(\beta_2|\beta_1)=0\\
  &&\bar P_2(\beta_2^c|\beta_1^c)=0  
\end{eqnarray}
where $\beta_1^c$ is driven by (\ref{Pg1}) .
\bigskip
We recall that in addition to (\ref{Pg1}) we have by conjugacy the
validity of
\begin{equation}\label{cPg1}
\bar P_g(\beta_1|\beta_1^c)=0 ~. 
\end{equation}

In the Appendix the following result is proven:

\bigskip

If $g(\beta_1,\beta_1^c)$ is irreducible
and
for every $\beta_1$
there exists a $\beta_1^c$ solution of
$g(\beta_1,\beta_1^c)=\bar g(\beta_1^c,\beta_1)=0$, then there exist
$\beta_1$, with $|\beta_1|$ as small as we like, such that
$\beta_1^c=\bar\beta_1$ is solution of the previous system.

\bigskip

A point on $V_0$ is given by a compatible pair of
$\beta_1,\beta_1^c$ i.e. a solution of (\ref{Pg1}) and (\ref{cPg1})
and by the Puiseux series \cite{whitney,nowak}
\begin{eqnarray}\label{1dimsheet}
  &&\beta_2 = a_1 \beta_1^{\frac{1}{m}}e^{\frac{2\pi i H}{m}}+
  a_2 \beta_1^{\frac{2}{m}}e^{\frac{4\pi i H}{m}}+\dots\\
  &&\beta_2^c = \bar a_1 (\beta_1^c)^{\frac{1}{m}}e^{\frac{2\pi i K}{m}}+
  \bar a_2 (\beta_1^c)^{\frac{2}{m}}e^{\frac{4\pi i K}{m}}+\dots
\end{eqnarray}
where the integers $H,K$ characterize the branch.
We know by conjugacy
that to the point $\beta_1,\beta_1^c,\beta_2,\beta_2^c$ there
correspond the point
$\bar\beta_1^c,\bar\beta_1,\bar\beta_2^c,\bar\beta_2$ which due to the
irreducibility assumption has to belong to the above branch.

The values of $\beta_2,\beta_2^c$ belonging to such a branch,
relative to a given compatible pair $\beta_1,\beta_1^c$ are
\begin{eqnarray}
  &&\beta_2 = a_1 \beta_1^{\frac{1}{m}} e^{\frac{2\pi i H}{m}}e^{\frac{2\pi i  nN}{m}}
  +a_2 \beta_1^{\frac{2}{m}} e^{\frac{4\pi i H}{m}}
  e^{\frac{4\pi i nN}{m}}+\dots\label{b2b1b2cb1cI}\\
  &&\beta_2^c = \bar a_1 (\beta_1^c)^{\frac{1}{m}}
  e^{\frac{2\pi i K}{m}e^{\frac{2\pi i nN}{m}}}+
  \bar a_2 (\beta_1^c)^{\frac{2}{m}}e^{\frac{4\pi i K}{m}}
  e^{\frac{4\pi i nN}{m}} +\dots\label{b2b1b2cb1cII}
\end{eqnarray}
where $n$ is the order of the $g$.  This is due to the fact that under
$\beta_1 \rightarrow \beta_1 e^{2\pi i n N}$ both $\beta_1$ and
$\beta_1^c$ are unchanged.  By conjugacy we must have that
substituting in (\ref{1dimsheet}), $\beta_1\rightarrow \bar\beta_1^c$
$\beta_1^c\rightarrow\bar\beta_1$ we find for some $N$, $\bar
\beta_2^c$ i.e.
\begin{equation}
  a_1(\bar\beta_1^c)^\frac{1}{m}e^{\frac{2\pi i H}{m}}e^{\frac{2\pi i nN}{m}}=
  \overline{(\bar a_1(\beta_1^c)^\frac{1}{m} e^\frac{2\pi i K }{m})}
\end{equation}  
i.e.
\begin{equation}\label{KHequation}
H+K+nN= m Z~.
\end{equation}  

Thus $H+K$ has to belong to the $(m,n)$ ideal. We ask now, whether
under such a restriction on $H+K$ we find a s.c. point on $V_0$ i.e.
$\beta_2^c=\bar\beta_2$ with $\beta_1^c=\bar\beta_1$. From
(\ref{b2b1b2cb1cI},\ref{b2b1b2cb1cII}) the problem is to find an $N_1$
such that
\begin{equation}
H+K+2nN_1= mZ_1~.
\end{equation}

Combining with the above
\begin{equation}\label{2N1equation}
n(2N_1-N)= mZ_2~.
\end{equation}
This equation in which $N_1$ and $Z_2$ are free is easily solved for
$m$ odd and when $m$ is even if also $N$ is even. To deal with the
case $m$ even and $N$ odd, we go over to the sheet obtained
by rotating $\beta_1$ by $e^{i\pi n}$ as done in the Appendix for even $n$.
Then eqs.(\ref{b2b1b2cb1cI},\ref{b2b1b2cb1cII}) become
\begin{eqnarray}\label{b2b1b2cb1cnew}
  &&\beta_2 = a_1 \beta_1^{\frac{1}{m}} e^{\frac{2\pi i H}{m}}
  e^{\frac{i\pi n}{m}}e^{\frac{2\pi i  nN}{m}}+
  a_2 \beta_1^{\frac{2}{m}} e^{\frac{4\pi i H}{m}}e^{\frac{2\pi i n}{m}}
  e^{\frac{4\pi i  nN}{m}}
  +\dots\\
  &&\beta_2^c = \bar a_1 (\beta_1^c)^{\frac{1}{m}}
  e^{\frac{2\pi i K}{m}e^{\frac{i\pi n}{m}}}e^{\frac{2\pi i nN}{m}}+
  \bar a_2 (\beta_1^c)^{\frac{2}{m}}e^{\frac{4\pi i K}{m}}e^{\frac{2i\pi n}{m}}
  e^{\frac{4\pi i nN}{m}} +\dots
\end{eqnarray}
which is like increasing $H$ and $K$ by $\frac{n}{2}$. Then
eq.(\ref{KHequation}) becomes 
\begin{equation}
H+K+n+nN= m Z 
\end{equation}  
which makes the new $N$ even ad thus eq.(\ref{2N1equation})
now soluble.

~

~

2. $g_j\equiv0$.

The variety is given by a branch of the W-type variety \cite{whitney}
\begin{eqnarray}\label{2dimrepresentation}
&&P(\beta_2|\beta_1,\beta_1^c)=0\nonumber\\
&&P^c(\beta_2^c|\beta_1,\beta_1^c)=\bar P^c(\beta_2^c|\beta_1^c,\beta_1)=0
\end{eqnarray}
where the $P$ are irreducible polynomials. In the present case we have
no constraint on $\beta_1$ and $\beta_1^c$.  We set $\beta_1=v
x$, $\beta_1^c=\bar v x$, with $|v|=1$.

Notice that for $x={\rm real}$ we have $\beta_1^c=\bar\beta_1$. If for
some $v$, $q(\beta_2,x)\equiv P(\beta_2|v x,\bar v x)$ is irreducible
we shall see right below that applying techniques similar to those
developed in the Appendix we find a solution of system (\ref{SYS2}),
with $\beta_1^c=\bar\beta_1$ and $\beta_2^c=\bar\beta_2$.
 
We know that $P(\beta_2|\beta_1,\beta_1^c)$ is irreducible, but from
this it does not follow necessarily that there exists a $v$ for which
$q(\beta_2,x)$ is irreducible. The irreducibility of the subvariety
$V_v$ given by (\ref{SYS2}) with $\beta_1=vx$, $\beta_1^c=\bar v x$
is an independent assumption.

We come now to the proof of the existence of a s.c. solution. We have
the two equations
\begin{eqnarray}
&&P(\beta_2|v x,\bar v x)=0\nonumber\\
&&\bar P(\beta_2^c|\bar v x, v x)=0
\end{eqnarray}
and all solutions to the previous are given by the Puiseux series
\begin{eqnarray}
&&\beta_2=c_1x^{\frac{1}{m}}+c_2x^{\frac{2}{m}}+...\nonumber\\
&&\beta_2^c=\bar c_1x^{\frac{1}{m}}+\bar c_2x^{\frac{2}{m}}+...
\end{eqnarray}
As these equations are identically satisfied we can send
$x^\frac{1}{m}\rightarrow e^\frac{2i\pi}{m}x^\frac{1}{m}$ and still
have a solution.  If for $x$ there is a point $vx,\bar
vx,\beta_2,\beta_2^c$ on $V_v$, i.e. solutions of (\ref{SYS2}) at
$u=u^c=0$, then there exist $h$ and $k$ such that
\begin{eqnarray}\label{fxxequation}
&&\beta_2=c_1x^{\frac{1}{m}}e^{\frac{2i\pi
    h}{m}}+c_2 \big(x^{\frac{1}{m}}e^{\frac{2i\pi h}{m}}\big)^2+\dots\nonumber\\
&&\beta_2^c=\bar c_1x^{\frac{1}{m}}e^{\frac{2i\pi
    k}{m}}+\bar c_2 \big(x^{\frac{1}{m}}e^{\frac{2i\pi k}{m}}\big)^2+\dots
\end{eqnarray}
where here by $x^{\frac{1}{m}}$ the principal value is
understood.

Such formulae must satisfy the conjugacy condition, i.e. that if we
send $\beta_1$ in $\bar\beta_1^c$ (and $\beta_1^c$ in $\bar\beta_1$)
we must have as solutions $\bar\beta_2^c$ and $\bar\beta_2$.  This
means that for some $N$
\begin{eqnarray}
&&c_1 x^{\frac{1}{m}}e^{\frac{2i\pi h}{m}}e^{\frac{2i\pi N}{m}}
+c_2 \big(x^{\frac{1}{m}}e^{\frac{2i\pi h}{m}}e^{\frac{2i\pi N}{m}}\big)^2+\dots
\nonumber\\  
&&\bar c_1x^{\frac{1}{m}}e^{\frac{2i\pi k}{m}}e^{\frac{2i\pi N}{m}}
  +\bar c_2 \big(x^{\frac{1}{m}}e^{\frac{2i\pi k}{m}}e^{\frac{2i\pi N}{m}}\big)^2+\dots
\end{eqnarray}
give respectively $\bar\beta_2^c$ and $\bar\beta_2$ for real $x$.
I.e.
\begin{equation}
c_1 x^{\frac{1}{m}}e^{\frac{2i\pi h}{m}}e^{\frac{2i\pi N}{m}}
=\overline{\bar c_1x^{\frac{1}{m}}e^{\frac{2i\pi k}{m}}}
\end{equation}
giving
\begin{equation}
h+k+N=mZ~.
\end{equation}
Given such a constraint on $h+k$ we look for a s.c. solution, i.e. for
an $N_1$ such that
\begin{equation}
c_1 x^{\frac{1}{m}}e^{\frac{2i\pi h}{m}}e^{\frac{2i\pi N_1}{m}}
=\overline{\bar c_1x^{\frac{1}{m}}e^{\frac{2i\pi k}{m}}e^{\frac{2i\pi N_1}{m}}}
\end{equation}
requiring
\begin{equation}
2N_1+h+k=2N_1-N+mZ = mZ_1~.
\end{equation}
For $m$ odd or $m$ even and $N$ even this equation is always soluble.
If $m$ is even and $N$ odd we work in (\ref{fxxequation}) on the real
negative $x$ axis as done in the Appendix. 
\begin{eqnarray}
&&\beta_2=c_1x'^{\frac{1}{m}}e^{\frac{2i\pi h}{n}}e^{\frac{i\pi}{m}}+\dots\\
&&\beta_2^c=\bar c_1x'^{\frac{1}{m}}e^{\frac{2i\pi k}{n}}e^{\frac{i\pi}{m}}+\dots
\end{eqnarray}
which amounts to increasing $h$ and $k$ by $1/2$. Then the conjugacy
condition imposes
\begin{equation}
h+k+1+N=mZ
\end{equation}
and now $N$ is even, which allows the above equation to be solved.

\section{Conclusions}\label{conclusions}

In several developments and applications of Liouville theory one needs
to exploit the real analytic nature of the dependence of the accessory
parameters on the moduli of the problem. Real analyticity is a
necessary requirement to extract e.g. the Polyakov relation
\cite{CMS1,CMS2,ZT,TZ} and its extension to the moduli space
\cite{torusblocks}.

Kra \cite{kra} proved using fuchsian mapping techniques that, for
parabolic singularities (punctures) and finite order singularities,
the accessory parameters are real analytic functions of the
moduli. Such a technique is not available in the case of general
elliptic singularities. On the other hand in most of applications one
deals with general elliptic singularities.

In the case of one independent accessory parameter like the sphere
topology with four sources or the torus with one source it was proven
\cite{torusIII,highergenus} that the accessory parameters are
real analytic functions except for a zero measure set in the moduli
space. A weaker result (real analyticity on an everywhere dense set)
had been proven in \cite{CMS1,CMS2}.

In section \ref{oneparameter} we give a simplified version of the
proof of such a.e. real analyticity.  Such a result follows only from
the general analytic properties of the solutions of the auxiliary
equation and the uniqueness theorem for the solution of Liouville
equation \cite{picard,poincare,lichtenstein,troyanov,existence}.

One naturally asks whether the general analytic properties of the
solutions of the auxiliary equation are again sufficient to establish
the real analyticity of the accessory parameter when the independent
parameters are more than one in number.

Thus in section \ref{twoaccessory} we considered the extension to the
case of two independent accessory parameters. Typical examples are the
sphere the five sources and the torus with two sources.

Again the aim is to extract all possible information from the general
analytic properties of the solutions of the auxiliary equation. The
main result obtained in section \ref{twoaccessory} is that, using the
results obtained in the one-parameter case,  one can
project the problem on a two dimensional (complex) plane; however this
is not enough for the general proof of real analyticity.
For this reason we examined in subsection \ref{irreduciblecase} the
irreducible case.

We have two possibilities where the dimension of the variety is $1$
or $2$.
For dimension $1$ irreducibility of $V_0$ is sufficient for reaching
the final result i.e. the proof of the real analytic dependence of the
two accessory parameters on the moduli of the problem, e.g. the
position of the sources and/or the moduli of the higher genus surface.

For dimension $2$ we need the irreducibility of the variety $V_v$.
After that again one proves the real analytic nature of the
dependence of the accessory parameters. 

Obviously the irreducibility of the manifolds $V_0$ or $V_v$ should be
proven by poking more deeply into the consequences of the auxiliary
equation (\ref{auxiliaryequation}) or by other procedures. We recall
that irreducibility of a variety is equivalent to the connectedness
property of the related manifold.

The question of the dependence of the accessory parameters on the
moduli, like the position of the sources, is not a trivial one when we
have more than one accessory parameter.  One could think to
extract the nature of the dependence from the constructive procedure
for finding the solution of the Liouville equation (\ref{liouville})
but this is not simple. The reason is that while the dependence of the
real field $\phi(z)$ on $z$ can be easily shown to be $C^\infty$ and
actually real analytic except at the sources, its dependence on the
positions of the sources is highly non trivial. The constructive proof
of the field $\phi$ goes through an iterative \cite{picard,poincare}
or minimization \cite{lichtenstein,troyanov,existence} procedure where
is not easy to follow the nature of the dependence on the position of
the sources.

In the present investigation we exploited only the very general
analytic properties of the solutions of the auxiliary equations. For
two or more accessory parameters such properties do not appear to be
sufficient for proving the real analyticity of the dependence of the
accessory parameters on the moduli on the problem and probably one has
to develop more deeply the consequences of the auxiliary
equation. Proving the irreducibility of the variety $V_0$ and $V_v$
would solve the problem.  On the other hand it is remarkable that
the general analytic properties of the solutions of the auxiliary
equation are sufficient for providing the complete proof of the real
analytic nature of the accessory parameter when we have a single
accessory parameter.

\section*{Appendix}

We prove here the following result:

\bigskip

Lemma: If $g(\beta_1,\beta_1^c)$ is irreducible in
$\Delta=\{|\beta_1|<\eta,|\beta_1^c|<\eta\}$  and
for every $\beta_1$  in $\Delta$ there exists a $\beta_1^c$ solution of
$g(\beta_1,\beta_1^c)=\bar g(\beta_1^c,\beta_1)=0$, then there exist
$\beta_1$, with $|\beta_1|$ as small as we like, such that
$\beta_1^c=\bar\beta_1$ is solution of the previous system.

\bigskip

Proof: We can always work with $g(\beta_1,\beta_1^c)$ in regular form.
Due to irreducibility all solutions of $g(\beta_1,\beta_1^c)=0$ are
given by the Puiseux series \cite{whitney,nowak} 
\begin{equation}\label{b1cb1}
\beta_1^c = B_h \beta_1^\frac{h}{n}+B_{h+1} \beta_1^\frac{h+1}{n}+\dots
\end{equation}
where $n$ is the order of the W-polynomial associated to $g$,
$\beta_1^\frac{1}{n}$ is one choice for the $n$-th root and thus we
have $n$ solutions. 
Similarly all solutions of  $\bar g(\beta_1^c,\beta_1)=0$ are
\begin{equation}\label{b1b1c}
\beta_1 = \bar B_h (\beta_1^c)^\frac{h}{n}+
\bar B_{h+1} (\beta_1^c)^\frac{h+1}{n}+\dots
\end{equation}
Having a common solution implies that for some $l$ and $s$ we have
\begin{equation}
\beta_1^c = B_h e^\frac{2\pi i h l}{n}\beta_1^\frac{h}{n}
+B_{h+1}e^\frac{2\pi i(h+1)l}{n}  \beta_1^\frac{h+1}{n}+\dots  
\end{equation}
\begin{equation}
\beta_1 = \bar B_h e^\frac{2\pi i h s}{n}(\beta_1^c)^\frac{h}{n}
+\bar B_{h+1}e^\frac{2\pi i (h+1) s}{n}(\beta_1^c)^\frac{h+1}{n}+\dots  
\end{equation}
where by $\beta_1^\frac{1}{n}$ we understood the principal value.
Consistency for small $\beta_1$ implies 
\begin{equation}
\beta_1 \beta_1^c= B_h \bar B_h e^{\frac{2\pi i h (l+s)}{n}}
(\beta_1 \beta_1^c)^\frac{h}{n}
\end{equation}
i.e. $h=n$ and $ B_h \bar B_h=1$. After multiplying $\beta_1$ by a
$v$, $|v|=1$ and $\beta_1^c$ by $\bar v$ we can rewrite the
equations (\ref{b1cb1},\ref{b1b1c}) as
\begin{eqnarray}
&&\beta_1^c = \beta_1(1+c_1\beta_1^\frac{1}{n}+
  c_2\beta_1^\frac{2}{n}+\dots\nonumber\\
&&\beta_1 = \beta_1^c(1+\bar c_1(\beta_1^c)^\frac{1}{n}+
\bar c_2(\beta_1^c)^\frac{2}{n}+\dots
\end{eqnarray}
and we have two choices for $v$ differing by sign.
Having a common solution now means, that for some $h$ and $k$ we have
\begin{eqnarray}\label{commonsolution}
&&\beta_1^c = \beta_1[1+c_1\beta_1^\frac{1}{n} e^\frac{2\pi i h}{n}+ c_2
(\beta_1^\frac{1}{n} e^\frac{2\pi i h}{n})^2+ \dots]\nonumber\\
&&\beta_1  = \beta_1^c [1+\bar c_1(\beta_1^c)^\frac{1}{n} e^\frac{2\pi i
  k}{n}+ \bar c_2
((\beta_1^c)^\frac{1}{n} e^\frac{2\pi i k}{n})^2+ \dots]
\end{eqnarray}
which implies
\begin{equation}
-c_1e^\frac{2\pi i h}{n}  =\bar c_1e^\frac{2\pi i k}{n}  
\end{equation}
i.e. for the argument $\varphi$ of $c_1$
\begin{equation}
\varphi+\frac{\pi}{2}=\frac{\pi}{n}(k-h)+\pi M,~~~~~~M=0,1.
\end{equation}
In order to reach the s.c. solution 
we go around the origin in $\beta_1$, $N_1$ times as to have
\begin{equation}
\overline{c_1 e^\frac{2\pi i (h+N_1)}{n}}=
\bar c_1 e^\frac{2\pi i (k+N_1)}{n}  
\end{equation}
i.e.
\begin{equation}\label{N1equation}
2N_1 = -(h+k)+n Z,~~~~~~~~Z={\rm integer}~. 
\end{equation}
We distinguish the two cases $n={\rm even}$ and $n={\rm odd}$~.

1) $n={\rm even}$.

For $h+k$ even this equation is always soluble by $2N_1=-(h+k)$, $Z=0$.

For $h+k$ even we have also the solution $2N_1=-(h+k)+n$ for $Z=1$,
which is obtained from the previous by $N_1\rightarrow N_1+n/2$.

For $h+k$ odd we perform the s.c. transformation
$\beta_1=e^{i\alpha}\beta_1',\beta_1^c=e^{-i\alpha}{\beta_1^c}'$ with
$\alpha=\pi$. The equations become renaming $\beta'_1,{\beta_1^c}'$
again as $\beta_1,\beta_1^c$
\begin{equation}\label{tonegative}
  \beta_1^c=\beta_1[1+c_1 e^{\frac{i\pi}{n}}
    \beta^\frac{1}{n}e^{\frac{2\pi i h'}{n}}+\dots]
\end{equation}
\begin{equation}
\beta_1=\beta_1^c[1+\bar c_1 e^{-\frac{i\pi}{n}}(\beta_1^c)^\frac{1}{n}
e^{\frac{2\pi ik'}{n}}+\dots]
\end{equation}
where we must allow for different $k',h'$. We have now the
compatibility restriction
\begin{equation}
-c_1 e^{\frac{i\pi}{n}} e^\frac{2i\pi h'}{n}  = \bar
c_1 e^{-\frac{i\pi}{n}}
e^\frac{2i\pi k'}{n}
\end{equation}
i.e.
\begin{equation}
  2\varphi+\pi +\pi\frac{(2h'+1)}{n}=
  \pi\frac{(2k'-1)}{n}+2 M\pi
\end{equation}
being $M$ an integer, i.e.
\begin{equation}
\varphi = \pi \big(-\frac{1}{2}+M+
  \pi\frac{(k'-h'-1)}{n}\big)
\end{equation}
while we had before
\begin{equation}
\varphi=\pi\big(-\frac{1}{2}+M+\frac{k-h}{n}\big)~.
\end{equation}
The s.c. requirement becomes now
\begin{equation}
\overline{e^{\frac{2\pi i (h'+N_1)}{n}}}=e^{\frac{2\pi i(k'+N_1)}{n}}
\end{equation}
i.e.
\begin{equation}
-\frac{(h'+N_1)}{n}=\frac{(k'+N_1)}{n}+Z
\end{equation}
which having changed the parity of $h+k$ is now soluble.

Again we have two solution, one with $N_1$ and the other with
$N_1\rightarrow N_1+n/2$.

As we shall see below for $n$ even, which we are dealing with, the
case $h+k$ even will give rise to s.c. solutions near the positive
real axis of $\beta_1$ while the case $h+k$ odd will give rise to
s.c. solutions near the negative real axis.

\bigskip

2) $n={\rm odd}$.

For $h+k$ even eq.(\ref{N1equation}) can be solved by $2N_1=-(h+k)$,
$Z=0$.

For $h+k$ odd  eq.(\ref{N1equation}) can be solved by
$2N_1=-(h+k)+n$, $Z=1$.

\bigskip

Summing up, renaming the $c_k$ we have that our solution obeys
\begin{eqnarray}
&&\beta_1^c=\beta_1[1+c_1 \beta^\frac{1}{n}+ c_2\beta^\frac{2}{n}+
    \dots] \label{betacbeta}\\
&&\beta_1=\beta_1^c[1+\bar c_1 (\beta_1^c)^\frac{1}{n}
+ \bar c_2(\beta_1^c)^\frac{2}{n}+ \dots]~.\label{betabetac}
\end{eqnarray}

\bigskip

In order to find a s.c. solution we solve in $\alpha(\rho)$ the
above equation with $\beta_1=\rho e^{i\alpha(\rho)}$ and
$\beta_1^c=\rho_ce^{-i\alpha(\rho)}$.  The variable $\alpha$ is
determined by the vanishing of the real function
\begin{equation}\label{alphaeq}
  i\bigg(e^{2i\alpha}(1+c_1\rho^{\frac{1}{n}} e^{i\frac{\alpha}
    {n}}
  +c_2 \rho^{\frac{2}{n}} e^{i\frac{2\alpha}{n}}+ \dots)-
    e^{-2i\alpha}(1+\bar c_1\rho^{\frac{1}{n}} e^{-i\frac{\alpha}
    {n}}
+\bar c_2 \rho^{\frac{2}{n}} e^{-i\frac{2\alpha}{n}}+ \dots)\bigg)
\end{equation}
which is soluble due to the implicit real function theorem.  Due to
the s.c. structure of our problem and using one of the two choices for
$v$ we have also
\begin{equation}\label{alphaeqcong}
\rho e^{i\alpha}= \rho_c e^{-i\alpha}(1+\bar c_1\rho_c^{\frac{1}{n}}
e^{-i\frac{\alpha}{n}}
+\bar c_2 \rho^{\frac{2}{n}} e^{-i\frac{2\alpha}{n}}+ \dots) ~.
\end{equation}
Taking the complex conjugate of eq.(\ref{alphaeqcong}) and multiplying
it by  eq.(\ref{betacbeta}) we obtain
\begin{equation}
\rho_c^2 (1+c_1\rho_c^{\frac{1}{n}}
e^{i\frac{\alpha}{n}}+ +c_2 \rho_c^{\frac{2}{n}}
e^{i\frac{2\alpha}{n}}+ \dots)=
\rho^2 (1+c_1\rho^{\frac{1}{n}}
e^{i\frac{\alpha}{n}}+ c_2 \rho^{\frac{2}{n}}
e^{i\frac{2\alpha}{n}}+ \dots)
\end{equation}
with the unique solution for small $\rho$, $\rho_c = \rho$. Thus we
constructed a s.c. solution to $g(\beta_1,\beta_1^c)=\bar
g(\beta_1^c,\beta_1)=0$
with ${\rm Re}\beta_1>0$ i.e. near the positive real axis.

For even $n$ we have in addition to the solution with $N_1$ the
solution with $N_1+n/2$ which gives the equations
\begin{equation}\label{alternative}
\beta_1^c=\beta_1[1-c_1 \beta_1^\frac{1}{n}+ c_2\beta_1^\frac{2}{n}- \dots]
\end{equation}
\begin{equation}
\beta_1=\beta_1^c[1-\bar c_1 (\beta_1^c)^\frac{1}{n}
+ \bar c_2(\beta_1^c)^\frac{2}{n}- \dots]
\end{equation}
and thus an other s.c. solution which according to the discussion
given above has \break ${\rm Re}\beta_1>0$, for $h+k$ even and 
${\rm Re}\beta_1<0$ for $h+k$ odd.

For odd $n$ we can rotate by the integer number of times
$\beta_1 \rightarrow \beta_1 e^{i(n+1)\pi}= -\beta_1 e^{i\pi n}$
and we have the solution
\begin{equation}\label{alternativenodd}
\beta_1^c=\beta_1[1-c_1 (-\beta_1)^\frac{1}{n}+ c_2(-\beta_1)^\frac{2}{n}- \dots]
\end{equation}
\begin{equation}
\beta_1=\beta_1^c[1-\bar c_1 (-\beta_1^c)^\frac{1}{n}
+ \bar c_2(-\beta_1^c)^\frac{2}{n}- \dots]~.
\end{equation}
Define $\beta_1'=-\beta_1$, ${\beta_1^c}'=-\beta_1^c$. Given the
s.c. structure of the coefficients we can now construct a
s.c. solution ${\beta_1^c}'=\bar\beta_1'$ with ${\rm Re}\beta_1' >0$,
i.e.  ${\rm Re}\beta_1 <0$.

Summarizing for $n$ even we have either two s.c. solutions near the
positive real axis or two s.c.  solution near the negative real axis,
while for $n$ odd we have always one s.c. solution near the positive
real axis and one near the negative real axis.

\eject

\vfill



\begin{thebibliography}{99}

\bibitem{CMS1} L. Cantini, P. Menotti, D. Seminara, {\it  Proof of
  Polyakov conjecture for general elliptic singularities}, 
  Phys.Lett. B517 (2001) 203

\bibitem{CMS2} L. Cantini, P. Menotti, D. Seminara, {\it Liouville
  theory, accessory parameters and (2+1)-dimensional gravity},
  Nucl.Phys. B638 (2002) 203

\bibitem{ZT} P.G. Zograf and L.A. Takhtajan, {\it On Liouville equation,
  accessory parameters, and the geometry of Teichm\"uller space for
  Riemann surfaces of genus $0$}, Math. USSR Sbornik 60 (1988) 143; 
 {\it On uniformization of Riemann surfaces and the Weyl-Peterson metric on
 Teichm\"uller and Schottky spaces}. Math.USSR Sbornik Vol 60 (1988) 297

\bibitem{TZ} L. A. Takhtajan, P. G. Zograf, {\it Hyperbolic 2 spheres
  with conical singularities, accessory parameters and K\"ahler metrics
  on M(0,n)}, Trans. Am.  Math. Soc. 355 (2003) 1857

\bibitem{torusblocks} P. Menotti, {\it Torus classical conformal blocks},
Mod.Phys.Lett. A33 (2018) no.28, 1850166

\bibitem{kra} I. Kra, {\it Accessory parameters for punctured spheres}, 
Trans. Am.Math.Soc. 313 (1989) 589

\bibitem{torusIII} P. Menotti, {\it  Accessory parameters for
  Liouville theory on the torus}, JHEP 12 (2012) 001

\bibitem{highergenus} P. Menotti, {\it The Polyakov relation for the
  sphere and higher genus surfaces}, J.Phys. A49 (2016) no.19, 195203

\bibitem{picard} E. Picard, {\it De l'integration de l'equation $\Delta
  u=e^u$ sur une surface de Riemann ferme\`e}, Journal de Crelle, 130
 (1905) 243

\bibitem{poincare} H. Poincar\'e, {\it Les functions fuchsiennes et
  l'equation $\Delta u=e^u$}, J. Math. Pures Appl. t.4 (1898) 137

\bibitem{lichtenstein} L. Lichtenstein, {\it Integration der
  Differentialgleichung $\Delta_2u=ke^u$ auf geschlossenen Fl\"achen},
  Acta mathematica 40 (1915) 1

\bibitem{troyanov} M. Troyanov, {\it Prescribing curvature on compact
  surfaces with conical singularities}, Transaction of the American
  Mathematical Society, 324 (1991) 793

\bibitem{existence} P. Menotti, {\it On the solution of the Liouville equation},
J.Phys. A50 (2017) no.37, 375205

\bibitem{whitney} H. Whitney, {\it Complex analytic varieties}, Addison
  Wesley Publ. Co., Reading, Massachusetts Menlo Park, California, 
  London Don Mills, Ontario, 1972

\bibitem{gunningrossi} R. C. Gunning, H. Rossi, {\it Analytic
  functions of several complex variables}, Prentice-Hall
  Inc.Englewood Cliffs (1965)

\bibitem{bochnermartin} S. Bochner and W. Martin, {\it Several complex
  variables}, Princeton University Press (1948)

\bibitem{harris} J. Harris, {\it Algebraic Geometry}, Springer-Verlag,
New York, 1992
  
\bibitem{nowak} K.J. Nowak, {\it Some elementary proofs of Puiseux's
  theorems}, U.Jagellonicae Acta Mathematica XXXVIII (2000) 279
  
\end{thebibliography}
\end{document}